\documentclass[
    reprint,twocolumn,
    aps,prl,10pt,amsmath,amssymb,
    longbibliography,superscriptaddress,
    showpacs,preprintnumbers,
    nofootinbib,
]{revtex4-2}

\usepackage{xr-hyper}
\usepackage[colorlinks=true, linkcolor=blue, citecolor=blue, urlcolor=blue, filecolor=blue]{hyperref}
\usepackage{bm,bbm,bbold}
\usepackage{braket}
\usepackage{mathtools}
\usepackage{amsmath,amssymb,amsfonts}
\usepackage[dvipsnames]{xcolor}
\usepackage{graphicx}
\usepackage[section]{placeins}
\usepackage{multirow}
\usepackage{orcidlink}
\usepackage[normalem]{ulem}  
\usepackage{xspace}
\usepackage{dcolumn}

\begin{document}

\title{Opposite impact of thermal expansion and phonon anharmonicity on the phonon-limited resistivity of elemental metals from first principles}

\author{Ao Wang}
\affiliation{European Theoretical Spectroscopy Facility, Institute of Condensed Matter and Nanosciences, Universit\'e catholique de Louvain, Chemin des \'Etoiles 8, B-1348 Louvain-la-Neuve, Belgium.}%

\author{Junwen Yin}
\affiliation{Scientific Computing Department, Science and Technology Facilities Council, UK Research and Innovation, Daresbury Laboratory, Keckwick Lane, Daresbury, WA4 4AD, UK.}%
\affiliation{European Theoretical Spectroscopy Facility, Institute of Condensed Matter and Nanosciences, Universit\'e catholique de Louvain, Chemin des \'Etoiles 8, B-1348 Louvain-la-Neuve, Belgium.}%

\author{F\'elix Antoine Goudreault}
\affiliation{D\'epartement de Physique et Institut Courtois, Universit\'e de Montr\'eal, C. P. 6128, Succursale Centre-Ville, Montr\'eal, Qu\'ebec, H3C 3J7, Canada.}%

\author{Michel C\^ot\'e}
\affiliation{D\'epartement de Physique et Institut Courtois, Universit\'e de Montr\'eal, C. P. 6128, Succursale Centre-Ville, Montr\'eal, Qu\'ebec, H3C 3J7, Canada.}%

\author{Olle Hellman}
\affiliation{Department of Molecular Chemistry and Materials Science, Weizmann Institute of Science, Rehovoth 76100, Israel.}%

\author{Samuel Ponc\'e}
\email{samuel.ponce@uclouvain.be}
\affiliation{European Theoretical Spectroscopy Facility, Institute of Condensed Matter and Nanosciences, Universit\'e catholique de Louvain, Chemin des \'Etoiles 8, B-1348 Louvain-la-Neuve, Belgium.}%
\affiliation{WEL Research Institute, Avenue Pasteur 6, 1300 Wavre, Belgium.}%

\date{\today}

\begin{abstract}
Understanding electrical resistivity in metals remains a central challenge in quantifying charge transport at finite temperature.
Current first-principles calculations based on the Boltzmann transport equation often match experiments, yet they almost always neglect the effect of thermal expansion and phonon anharmonicity.
We show that both effects exert an opposite impact on electron–phonon coupling and on electrical resistivity.  
Thermal expansion enhances the coupling and leads to overestimation of resistivity, whereas anharmonic effects reduce it.
By explicitly incorporating both effects, we establish a more complete description of resistivity in elemental metals, demonstrated here for Pb, Nb, and Al. 
\end{abstract}

\maketitle

Electrical resistivity, one of the most important physical quantities in metals, has attracted sustained research interest over the past hundred years.
Since the observation of the linear temperature dependence of electrical resistivity in pristine metals near room temperature by early experiments~\cite{Jaeger1900,Powell1967}, a long-term development in the theory has occurred to describe it.
Based on the kinetic theory of gases, the Drude model~\cite{Drude1900,Drude_2_1900} was proposed to explain the linear relationship between voltage and current. 
However, the quantitative electrical resistivity was unpredictable as a result of the undetermined microscopic relaxation time.
With quantum physics established, Allen's model~\cite{Allen1978} was successful in quantitatively determining the electrical resistivity of several metals~\cite{Grimvall1981,Savrasov1996,Bauer1998,Grimvall1999} by connecting the resistivity to transport spectral function.
The development of first-principles codes to interpolate electron-phonon matrix elements~\cite{Giustino2007,Ponce2016,Gonze2016,Gonze2020,Zhou2021,Cepellotti2022,Lee2023,Marini2024} and solve the Boltzmann transport equation (BTE) iteratively considering mode-resolved electron-phonon interactions, allowed for the study of the electrical resistivity of pure metals and metallic compounds~\cite{Li2015,Gall2016,Tong2019,Giri2020,Alvarez2025}, mostly in agreement with experiments.

Most theoretical models and numerical calculations neglect the thermal expansion (TE), inconsistent with the underlying physics.
Recent studies have found that TE is crucial for the calculation of the temperature-induced renormalization of the gap~\cite{FranciscoLpez2019,Rubino2020}, phonon dispersion~\cite{Tang2024}, and phonon-phonon interactions~\cite{Tang2024,Zeng2024}, and should not be neglected.
Since it is anharmonicity that drive TE, its effect should also be included in the evaluation of electrical conductivity.
A recent study~\cite{Goudreault2025} showed that accounting for TE in Pb leads to a strong overestimation of resistivity at high temperature compared to the experiment
(+92\% at 610~K).
The authors tested the impact of several factors to account for this overestimation, including spin-orbit coupling (SOC)~\cite{Verstraete2008} and various exchange-correlation functionals, but did not resolve it. 

In this work, we show that including phonon anharmonicity~\cite{Wei2021} opposes the effect of lattice TE and that the two effects should always be added together.  
In practice, phonon anharmonicity can be included with the self-consistent phonon (SCP) method~\cite{Errea2014,Tadano2018} using the self-consistent $ab$-$initio$ lattice dynamical (\textsc{SCAILD}) package~\cite{Souvatzis2008}, temperature dependent effective potentials (TDEP) method~\cite{Hellman2013} and \textsc{TDEP} package~\cite{Knoop2024}, stochastic self-consistent harmonic approximation (SSCHA) method~\cite{Errea2014} and \textsc{SSCHA} package~\cite{Monacelli2021}, or the $ab$-$initio$ molecular dynamic (aiMD) method~\cite{Iftimie2005}. 
Previous works~\cite{Romero2015,Zhou2018,Yin2025} found that phonon anharmonicity is crucial to determining the carrier mobility of semiconductors but the electrical resistivity of metals has not been investigated yet. 

To include TE effects, we calculate the Helmholtz free energy $F(V,T)$ with both electron and phonon contributions. 
The electron part is composed of the clamped-ion energy at 0~K and the energy of the thermal electronic excitation. 
For the phonon contribution, we use the quasi-harmonic approximation (QHA)~\cite{Lazzeri1998,Carrier2007,Allen2019}.
We use the same computational parameters as Ref.~\cite{Goudreault2025}. 
We compute $F(V,T)$ from density functional theory (DFT) with \textsc{Quantum ESPRESSO}~\cite{Giannozzi2017,Giannozzi2020} using a third-order Birch-Murnaghan isothermal fit~\cite{Birch1947,Murnaghan1944}, which produces the temperature-dependent lattice parameter.
To include anharmonic effects, we use \textsc{TDEP}~\cite{Knoop2024} where the Hamiltonian is expanded to the second-order with respect to the atom displacements~\cite{Hellman2013}. 
We use stochastic sampling~\cite{Shulumba2017} in which the configurations with atomic displacements are generated by the canonical ensemble from the interatomic force constants (IFCs) and the effective IFCs are determined by linear least-squares solution with the displacements and forces. 
The IFCs and atomic displacements can therefore be iteratively calculated at different temperatures. 
Additional anharmonic computational details and convergence studies for Pb are provided in Sec.~S1 of the Supplemental Information (SI)~\cite{Wang2025} (see also references
~\cite{Nenno1960, Monkhorst1976, Pack1977, Bandyopadhyay1978, Baroni1987, Gonze1989, Vanderbilt1990, Perdew1996, Wang1998, Marzari1999, Hamann2013, DalCorso2014, Shulumba2017, Giannozzi2017, vanSetten2018, Macheda2018, Giannozzi2020, Knoop2024, Goudreault2025} therein).

The electrical resistivity $\rho_{\alpha\beta}$ is obtained by solving the linearized Boltzmann transport equation~\cite{Ponce2020}
\begin{equation}\label{eq:rho}
    \rho_{\alpha\beta}^{-1} = -\frac{e}{(2\pi)^3} \sum_n \int \frac{\rm{d}^3{\bf{k}}}{\Omega^{\rm BZ}} v_{n\mathbf{k}\alpha} \, \partial_{E_\beta} f_{n\mathbf{k}},
\end{equation}
where $e$ is the electron charge, $\alpha$ and $\beta$ Cartesian coordinates, $n\mathbf{k}$ is the band index and the wavevector of electrons,  $\Omega^{\rm{BZ}}$ is the volume of the first Brillouin zone, and $v_{n\mathbf{k}\alpha}$ is the electron velocity.
The $\partial_{E_\beta} f_{n\mathbf{k}}$ term in Eq.~\eqref{eq:rho} is the change of occupation due to the external electric field $E_\beta$. 
This term is obtained as the solution of~\cite{Ponce2018,Ponce2020,Claes2025}
\begin{multline}\label{eq:BTE}
\tau _{n\mathbf{k}}^{-1} \partial_{E_\beta}{f_{n\mathbf{k}}} = e \frac{\partial f_{n\mathbf{k}}^0}{\partial{\varepsilon_{n\mathbf{k}}}}{v_{n \mathbf{k}\beta}} + \frac{2\pi}{\hbar} \sum_{m\nu} \int \frac{\textrm{d}^3 \mathbf{q}}{\Omega^{\textrm{BZ}}} |g_{mn\nu}(\mathbf{k,q})|^2 \\
\times \Big[(1 + n_{\mathbf{q}\nu} - f_{n \mathbf{k}}^0) \delta( \varepsilon_{n\mathbf{k}} -  \varepsilon_{m\mathbf{k+q}} + \hbar {\omega _{\bf{q}\nu }}) \\
+ (n_{\mathbf{q}\nu} + f_{n\mathbf{k}}^0) \delta(\varepsilon_{n\mathbf{k}} - \varepsilon_{m\mathbf{k+q}} - \hbar {\omega _{\mathbf{q}\nu}}) \Big] \partial_{E_\beta}{f_{m\mathbf{k+q}}},
\end{multline}
where $\varepsilon_{n\mathbf{k}}$ are the eigenenergies, $f_{n \mathbf{k}}^0$ is the equilibrium Fermi-Dirac distribution for electrons, $n_{\mathbf{q}\nu}$ is the Bose-Einstein distribution for phonons, and $\omega_{\mathbf{q}\nu}$ is the frequency of the phonon with the wavevector $\mathbf{q}$ and the branch index $\nu$.
The $g_{mn\nu}(\mathbf{k,q})$ are the electron-phonon coupling (EPC) matrix element, which represents the probability amplitudes of the scattering of electron from state $n \mathbf{k}$ to $m\mathbf{k+q}$ by a phonon $\mathbf{q}\nu$.
We neglect the $g_{mn\nu}(\mathbf{k,q})$ for which $\omega_{\mathbf{q}\nu} < 5$~cm$^{-1}$.
In Eq.~\eqref{eq:BTE}, we have introduced the electron-phonon scattering rate as~\cite{Ponce2018}
\begin{multline}
\tau_{n\mathbf{k}}^{-1} \equiv \frac{2\pi}{\hbar} \sum_{m\nu} \int {\frac{ \textrm{d}^3\mathbf{q}}{ \Omega^{\rm{BZ}}}} | g_{mn\nu}(\mathbf{k,q})|^2 \\
\times \Big[ (1 - f_{m\mathbf{k+q}}^0 +  n_{\mathbf{q}\nu}) \delta( \varepsilon_{n\mathbf{k}} -  \varepsilon_{m\mathbf{k+q}}  - \hbar \omega_{\mathbf{q}\nu} )  \\
+ ( f_{m\mathbf{k+q}}^0 + n_{\mathbf{q}\nu}) \delta( \varepsilon_{n\mathbf{k}} - \varepsilon_{m\mathbf{k+q}} + \hbar \omega_{\mathbf{q}\nu}) \Big].
\end{multline}

We use \textsc{Quantum ESPRESSO}~\cite{Giannozzi2017,Giannozzi2020} and \textsc{TDEP}~\cite{Knoop2024} for the anharmonic IFC, together with \textsc{Wannier90}~\cite{Pizzi2020} and \textsc{EPW}~\cite{Giustino2007,Ponce2016,Lee2023} for the electrical resistivity calculations.
We use the recently developed \textsc{EPW}-\textsc{TDEP} interface~\cite{Yin2025} and SOC is used throughout for Pb. 
Importantly, although IFCs and $g_{mn\nu}(\mathbf{k,q})$ include temperature-dependent anharmonic effects, the perturbed potential is computed using the harmonic approximation.
Additional details and convergence can be found in Sec.~S2 of the SI~\cite{Wang2025}.

\begin{figure}[t]
    \centering
    \includegraphics[width=0.95\linewidth]{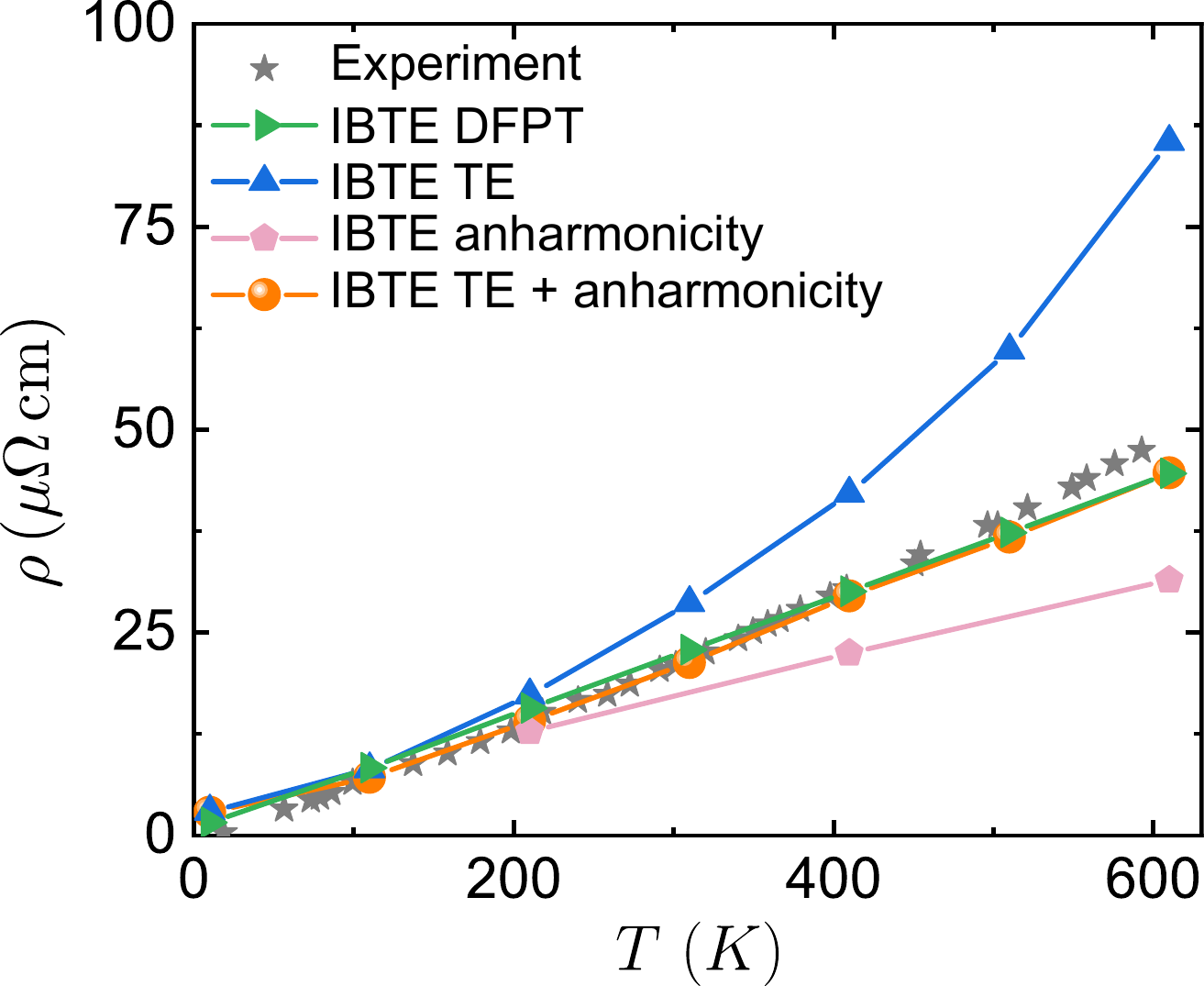}
    \caption{\label{fig:fig1}
    Electrical resistivity of Pb from 10 to 610~K using the iterative Boltzmann transport equation (IBTE). 
    The green and blue markers are computed without and with thermal expansion (TE), without considering the phonon anharmonicity, while the computation of pink and orange markers takes the phonon anharmonicity into account. 
    The experimental data are from Ref.~\cite{hellwege1982}.
    }
\end{figure}

\begin{figure*}[t]
    \centering
    \includegraphics[width=0.99\linewidth]{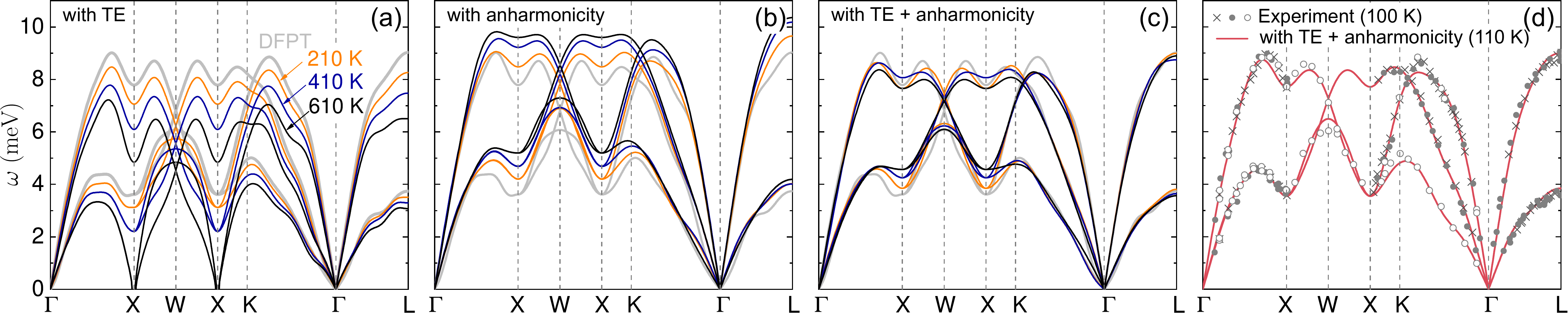}
    \caption{\label{fig:fig2}
    Phonon dispersion of Pb including (a) lattice thermal expansion (TE) effect where calculations are harmonic with a temperature-dependent lattice parameters, (b)
    without TE and with temperature-dependent phonon anharmonicity, and (c) with TE and phonon anharmonicity. 
    In all cases the gray line is the harmonic phonon dispersion computed with density functional perturbation theory (DFPT) without TE. 
    (d) Phonon dispersion of Pb calculated with TE and phonon anharmonicity at 110~K compared with experimental data at 100~K measured by neutron scattering from Ref.~\onlinecite{Brockhouse1962} ($\times$ markers) and from Ref.~\onlinecite{Brockhouse1961} (open and closed circles).
    }
\end{figure*}

The electrical resistivity of Pb from 10 to 610~K is presented in Fig.~\ref{fig:fig1}.
If we ignore the effects of phonon anharmonicity and lattice TE, the resistivity at 310~K is 22.8~$\mu\Omega\text{cm}$, close to the experimental value of 21.7~$\mu\Omega\text{cm}$~\cite{hellwege1982}. 
This result also closely reproduces a prior work~\cite{Goudreault2025}, see Fig.~S3 of the SI~\cite{Wang2025} for a comparison.
However, if the effect of lattice TE is included, the electrical resistivity increases greatly from 210 to 610~K, with an enhancement of 92\% at 610~K.
This overestimation is mostly attributed to phonon softening, especially at the $\mathbf{q}=\mathbf{X}$ point, as seen in Fig.~\ref{fig:fig2}(a).
In fact, the softening is so strong that the crystal becomes thermodynamically unstable at 610~K, which coincides with the melting point of Pb. 
Interestingly, the impact of TE on the electronic bandstructure is minimal, see Fig.~S4 of the SI~\cite{Wang2025}.

In contrast, we find that the impact of including phonon anharmonicity is opposite to including TE and leads to a phonon hardening with temperature, see 
 Fig.~\ref{fig:fig2}(b).
This effect in turn results in an underestimated electrical resistivity as seen in Fig.~\ref{fig:fig1} with pink pentagons.
Therefore, including both effects leads to the phonon frequencies reported in Fig.~\ref{fig:fig2}(c), which agree perfectly with experimental data~\cite{Brockhouse1962,Brockhouse1961} at the same temperature, see Fig.~\ref{fig:fig2}(d).
In addition, the resulting computed resistivity shown in orange in Fig.~\ref{fig:fig1} matches the experimental results well, even at higher temperature. 
Moreover, we note that including both effects or excluding both of them result in close transport properties. 
This suggests that the TE and phonon anharmonicity show an antagonistic impact on the resistivity of Pb.

\begin{figure*}[t]
    \centering
    \includegraphics[width=0.95\linewidth]{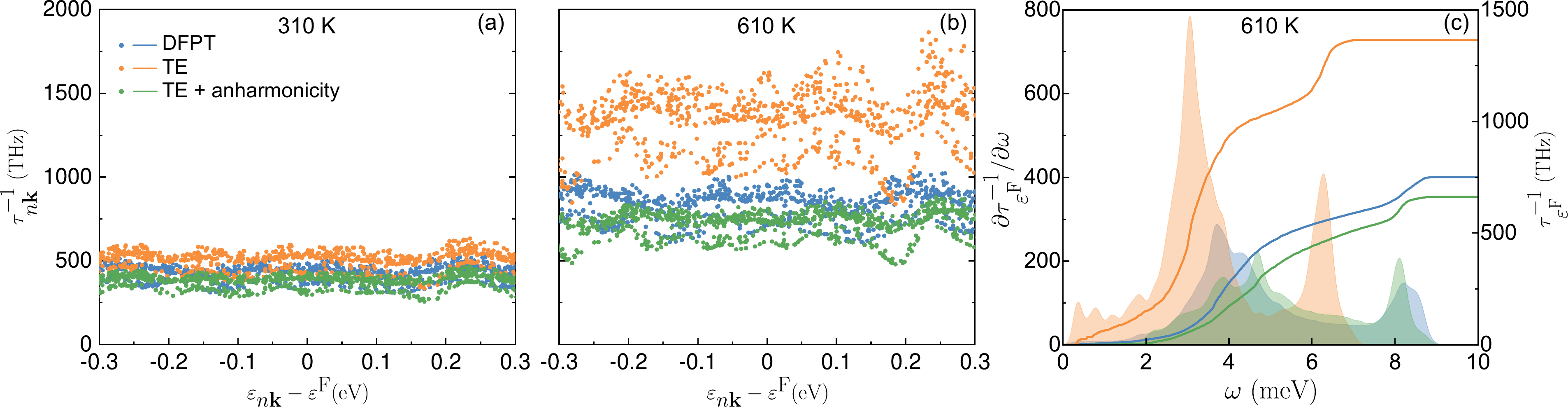}
    \caption{\label{fig:fig3}
    Electron scattering rates $\tau_{n\mathbf{k}}^{-1}$ of Pb at $\pm$ 0.3~eV around the Fermi energy $\varepsilon^{\rm F}$ at (a) 310~K and (b) 610~K. 
    (c) Spectral decomposition of the electron scattering rates averaged around the fermi energy as a function of phonon frequency at 610~K.
    In each subfigures, we show results without thermal expansion (TE) and without phonon anharmoncity (blue), with TE and without anharmonicity (orange), and with TE and anharmoncity (green).
    }
\end{figure*}

To understand the microscopic origin of this cancellation, we calculated the electron scattering rates $\tau_{n\mathbf{k}}^{-1}$ in Fig.~\ref{fig:fig3}. 
We show the results for carriers with energy $\pm$ 0.3~eV around the Fermi energy $\varepsilon^{\rm F}$ which contributes to electrical resistivity.
As shown in Fig.~\ref{fig:fig3}(a), we find that at 310~K the TE leads to a 20\% increase in scattering rates, which explains the 25\% increase in the electrical resistivity at 310~K in Fig.~\ref{fig:fig1}. 
With phonon anharmonicity further included, the scattering rates decrease and the values are then close to those computed without TE and anharmonicity.
This phenomenon is consistent with both the phonon dispersion and the electrical resistivity. 

When the temperature increases to 610~K, as shown in Fig.~\ref{fig:fig3}(b), the impact of TE and phonon anharmonicity on scattering rates is quite similar to the case at 310~K. 
The only difference is that, at higher temperatures, both the impact of the two properties on scattering rates becomes more pronounced.
To further investigate the electron-phonon scattering process, we also show the spectral decomposition of the electron scattering rates at 610~K in Fig.~\ref{fig:fig3}(c).
We find that when only TE is considered, the phonons with energies from 2 to 4~meV provide most of the scattering.
The contribution to electron scattering from phonons with energies smaller than 2~meV is not so strong, despite a soft mode at the $\mathbf{q}=\mathbf{X}$ point due to their 
smaller phonon density of states.
This analysis highlights the relationship between electron-phonon scattering and phonon softening.
When phonon anharmonicity is further included, the spectral decomposition becomes quite close to the case in which both finite-temperature effects are ignored. 
In addition, a slight increase in the phonon energy of around 4~meV of the spectral decomposition is found, which is consistent with the gradually eliminated softening phonons with temperatures shown in Fig.~\ref{fig:fig2}(c).

\begin{figure}[t]
    \centering
    \includegraphics[width=0.85\linewidth]{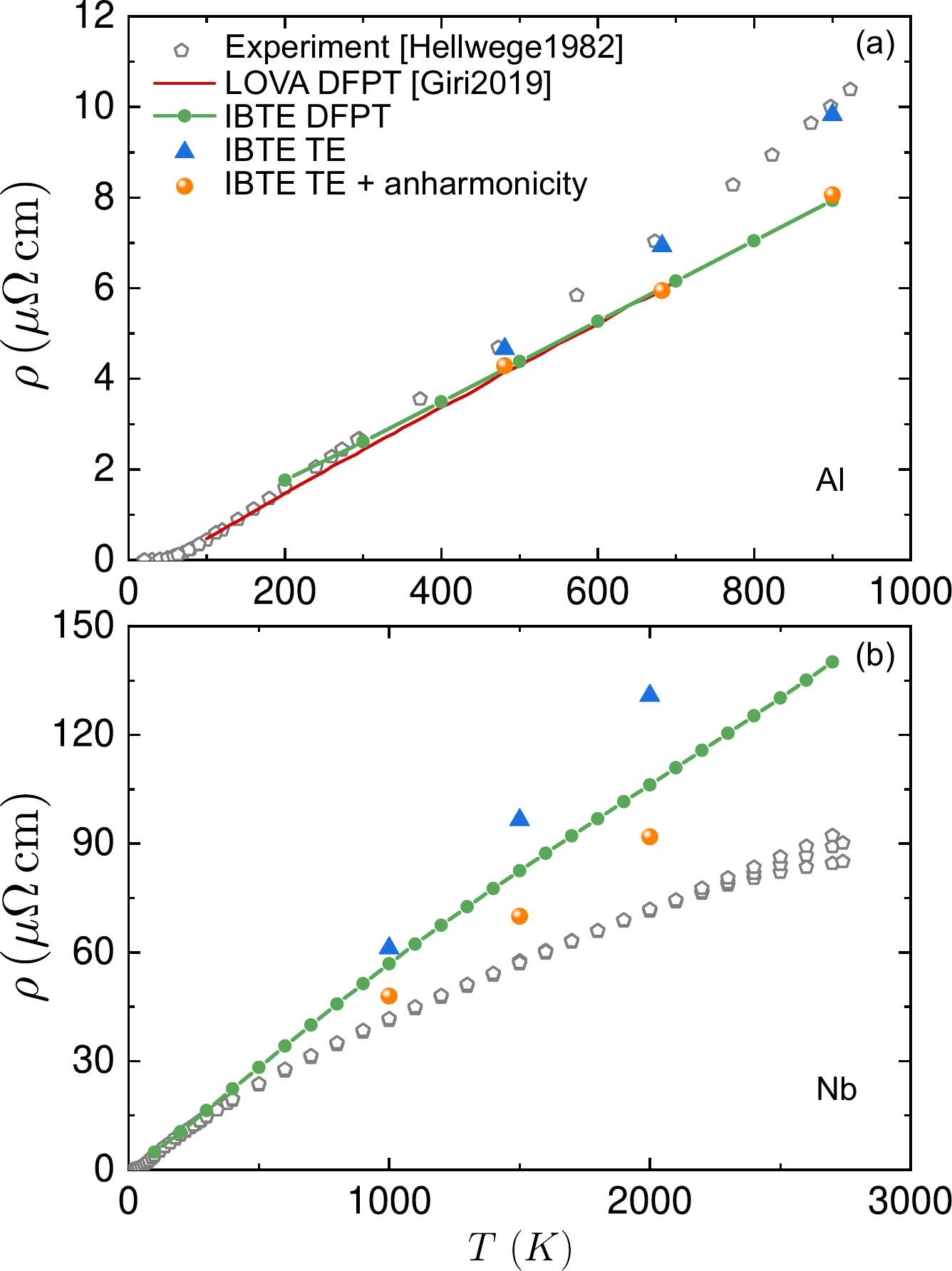}
    \caption{\label{fig:fig4}
    Electrical resistivity of (a) Al from 0 to 1000~K and (b) Nb from 0 to 3000~K using the iterative Boltzmann transport equation (IBTE). 
    The green and blue markers are computed without and with thermal expansion (TE), without considering the phonon anharmonicity, while the computation of orange markers takes the phonon anharmonicity into account. 
    The experimental data are from Ref.~\cite{hellwege1982} and the red line is obtained with the lowest order variational approximation (LOVA) from Ref.~\cite{giri2019}.}
\end{figure}

To assess the generality of this antagonistic effect between lattice TE and phonon anharmonicity,  we further investigated the electrical resistivity 
of aluminum and niobium.
Unlike Pb, Nb shows a Kohn anomaly related to its nesting Fermi surface~\cite{Powell1968,Tidholm2020}.
For Al and Nb we use the experimental temperature-dependent lattice constant~\cite{Bandyopadhyay1978,Nenno1960,Wang1998} and provide all computational 
details and convergence studies in Sec.~S3 of the SI~\cite{Wang2025}.
The calculated electrical resistivity matches quite well with experiments below 500~K in Al and Nb when both TE and anharmonicity are excluded, with the comparison shown in Fig.~\ref{fig:fig4}.
We find that the case of Al is very similar to Pb and that the impact of TE and phonon anharmonicity is opposite and of similar magnitude. 
We note that there is an underestimation between calculations and experiments above 500~K.
A similar underestimation was found in a previous theoretical study using the lowest order variational approximation (LOVA)~\cite{giri2019}, an approximated solution to the BTE.

Interestingly, we find that the case of Nb is different, and although the impact of TE and phonon anharmonicity is also opposite, they do not cancel each other. 
In this case, the effect of the phonon anharmonicity is stronger, bringing the results closer to experiment~\cite{hellwege1982}.
We attribute the remaining overestimation to saturation effects~\cite{Gunnarsson2003}.
The lack of cancellation effects in Nb as compared to Pb and Al can be attributed to the highly dispersive nature of phonon shifts with temperature. 
When the shifts due to thermal expansion or anharmonicity concentrate in different parts of the Brillouin zone, one can not expect a direct cancellation of the effects on resistivity. 
We can relate this to the distinctly different origin of the shifts caused by volume change and temperature change. 
A volume change corresponds to a rigid stretching of bonds whereas at increased temperature Nb sees a smearing of its Fermi surface that screens the nesting vectors.
At low temperature, flat sheets of the Fermi surface can be connected with a phonon wavevector which can be identified as the Kohn anomaly along $\Gamma-H$. 
At elevated temperature, the smearing of the Fermi surface removes the singular behavior and the phonon dispersions straighten~\cite{Tidholm2020}.
We speculate that this is a general behaviour in metals that have intricate Fermi surface geometry and/or bands in close proximity where temperature effects can bring them into contact with the Fermi surface and alter its topology~\cite{Yang2016}.

In summary, we investigated the effect of TE and phonon anharmonicity on electrical resistivity in Pb, Nb, and Al using first-principle calculations at different temperatures. 
We find that as temperature increases, TE leads to a larger resistivity due to stronger electron-phonon scattering resulting from phonon softening. 
However, phonon anharmonicity results in a smaller resistivity because of phonon hardening with temperature.
These two mechanisms show an antagonistic impact on the electrical resistivity of these simple metals. 
Only when both effects are considered can the theoretical results be matched with the experimental results.
In addition, both these mechanisms affect the spectral decomposition of the electron scattering, and with both effects considered, electrons for transport in Pb are mostly scattered by phonons with energy from 3 to 5 meV.

\begin{acknowledgments}
S. P. is a Research Associate of the Fonds de la Recherche Scientifique - FNRS.
This publication was supported by the Walloon Region in the strategic axe FRFS-WEL-T.
Computational resources have been provided by the EuroHPC JU award granting access to MareNostrum5 at Barcelona Supercomputing Center (BSC), Spain (Project ID: EHPC-EXT-2023E02-050), by the Consortium des \'Equipements de Calcul Intensif (C\'ECI), funded by the FRS-FNRS under Grant No. 2.5020.11, by the Tier-1 supercomputer of the Walloon Region (Lucia) with infrastructure funded by the Walloon Region under the grant agreement No. 1910247.
\end{acknowledgments}

\emph{Data Availability.} 
The data that support the findings of this article are openly  available~\cite{Wang2025_2}.

\bibliography{Bibliography}

\end{document}